\documentclass{elsart}

\usepackage{graphics}
\usepackage{graphicx}
\usepackage{epsfig}

\usepackage{amssymb}

\begin{document}

\begin{frontmatter}



\title{ESR investigation of the spin dynamics in (Gd$_{1-x}$Y$_{x}$)$_2$PdSi$_{3}$}


\author{J. Deisenhofer, H.-A. Krug von Nidda, and A. Loidl}
\address{Experimentalphysik V, EKM, Universit\"{a}t Augsburg,
86135 Augsburg, Germany}
\author{E.V. Sampathkumaran}
\address{Tata institute of Fundamental Research, Homi Bhabha Road, Mumbai-400 005, India}

\begin{abstract}
We report electron-spin resonance measurements in polycrystalline
samples of (Gd$_{1-x}$Y$_{x}$)$_2$PdSi$_{3}$. We observe the onset
of a broadening of the linewidth and of a decrease of the
resonance field at approximately twice the N\'{e}el temperature in
the paramagnetic state, this characteristic temperature coincides
with the electrical resistivity minimum. The high-temperature
behaviour of the linewidth is governed by a strong bottleneck
effect.
\end{abstract}

\end{frontmatter}

\section{Introduction}
\label{}

It is of general believe that due to the close $f$-shell Gd ions
do not show exotic magnetic behaviour in solid-state environment.
However, recently it has been reported that some Gd alloys exhibit
interesting phenomena including large magnetoresistance effects
\cite{Eremin01,Felser99,Ahn00,Mallik98a}. A
metal-to-insulator-like transition has been reported for
Gd$_5$Ge$_4$ \cite{Levin01}. In addition, a large number of the
Gd-based intermetallics reveal large negative magnetoresistance
(MR) effects arising from peculiar magnetic precursor effects
\cite{Mallik98}. Recently, the Gd transition-metal silicides and
germanides with 2:1:3 stoichiometry (AlB$_2$-type hexagonal
structure \cite{Kotsanidis90}) came into focus, as some of these
compounds even show a Kondo-like minimum in the temperature
dependence of the resistivity well above the magnetic ordering
temperatures \cite{Mallik98a,Majumdar00,Majumdar01}. In
polycrystalline Gd$_2$PdSi$_{3}$ ($T_N=21$ K), the compound of
interest in this article, a pronounced minimum occurs at about 45
K, which has been confirmed in single crystalline Gd$_2$PdSi$_3$
Saha \textit{et al.~}\cite{Saha99}. The minimum persists even when
diluting the Gd sublattice by Y in
(Gd$_{1-x}$Y$_{x}$)$_2$PdSi$_{3}$ \cite{Mallik98a}. Chaika
\textit{et al.~}investigated the electronic structure for
$R_2$PdSi$_3$ with $R$=La,Ce,Gd and Tb by photoemission and
band-structure calculations \cite{Chaika01} and found that in
Gd$_2$PdSi$_3$ the Gd 4\textit{f} level  is far below the Fermi
level and hence should not significantly hybridize with the band
states. In order to shed some light on the unusual magnetic and
electronic properties of this compound we systematically
investigated the spin dynamics of
(Gd$_{1-x}$Y$_{x}$)$_2$PdSi$_{3}$ by electron-spin resonance
(ESR).

\section{Experimental details and results}
\label{}

The samples were prepared by arc melting and are taken from the
same batch as those used in previous investigations by Mallik
\textit{et al.~}\cite{Mallik98a}. ESR measurements were performed
in a Bruker ELEXSYS E500 CW-spectrometer at X-band frequencies
($\nu \approx$ 9.47 GHz) equipped with a continuous He-gas-flow
cryostat in the temperature region $4.2<T< 300$ K. The
polycrystalline samples were powdered, placed into quartz tubes
and fixed with paraffin. ESR detects the power $P$ absorbed by the
sample from the transverse magnetic microwave field as a function
of the static magnetic field $H$. The signal-to-noise ratio of the
spectra is improved by recording the derivative $dP/dH$ with
lock-in technique.

\subsection{ESR spectra and intensity}

\begin{figure}[ht]
\centering
\includegraphics[width=80mm,clip,angle=-90]{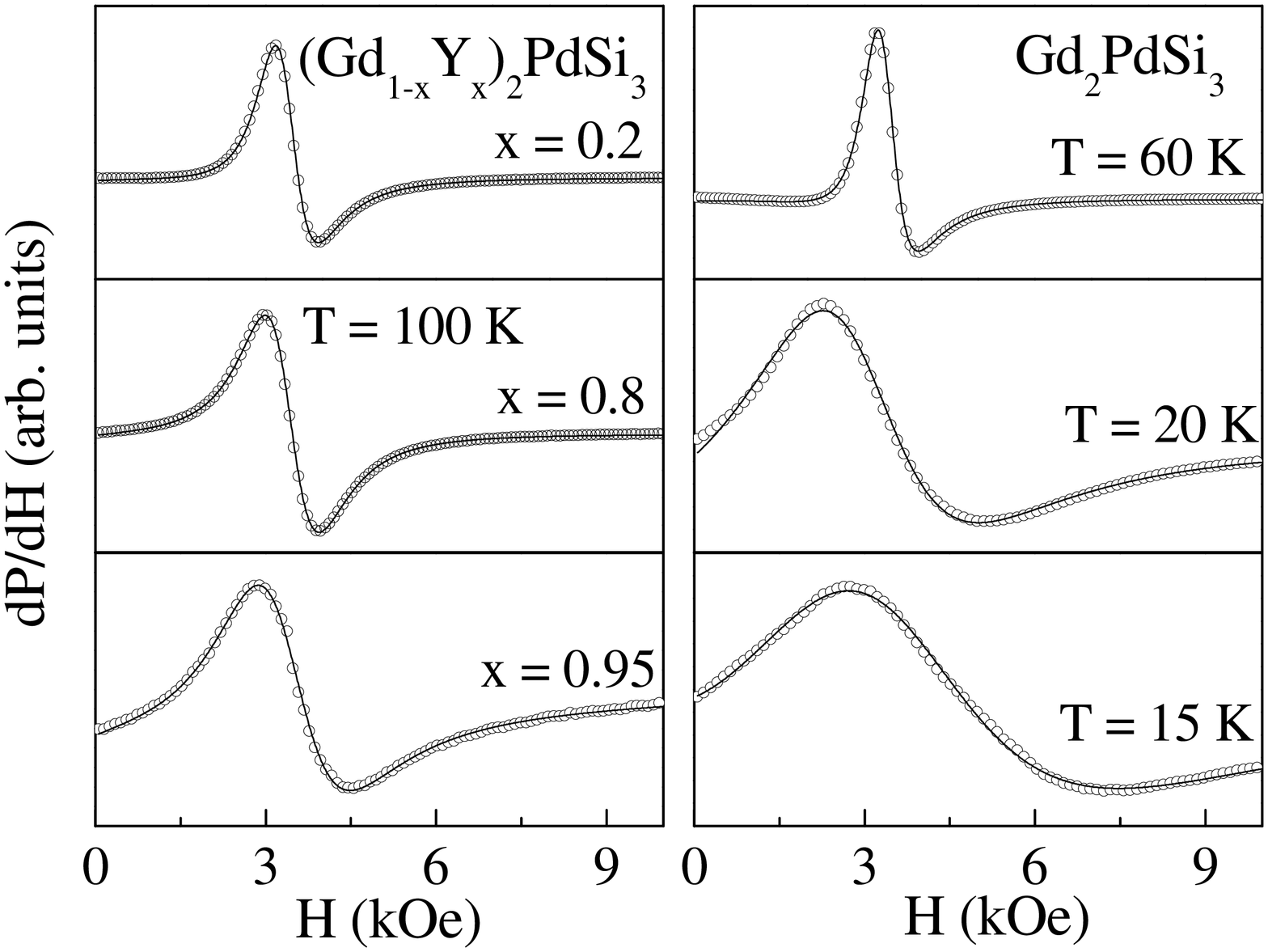}
\vspace{2mm} \caption[]{\label{spectra} ESR spectra of
(Gd$_{\mathrm{1-x}}$Y$_{\mathrm{x}})_2$PdSi$_3$. Left column:
various Y concentrations $x$ at $T = 100$ K. Right column:
temperature evolution of the ESR spectrum for $x = 0$. Solid lines
represent the fits using the Dysonian line shape, Eq. (1).}
\end{figure}

Typical ESR spectra are presented in Fig.~\ref{spectra} and
illustrate the evolution of the Gd resonance with Y concentration
$x$ (left column) and temperature $T$ (right column).
As expected, no ESR signal could be observed for $x=1$. For $x<1$,
within the whole paramagnetic regime the spectrum consists of a
broad, exchange narrowed resonance line, which is well fitted by a
Dysonian line shape \cite{Feher55}. As in the present compounds
the linewidth $\Delta H$ is of the same order of magnitude as the
resonance field $H_{res}$, both circular components of the
exciting linearly polarized microwave field have to be taken into
account. Therefore the resonance at the reversed magnetic field
$-H_{res}$ has to be included into the fit formula for the ESR
signal, given by
\begin{equation}
\frac{dP}{dH} \propto \frac{d}{dH}\left\{\frac{\Delta H + \alpha
(H-H_{res})}{(H-H_{res})^2 + \Delta H^2} + \frac{\Delta H + \alpha
(H+H_{res})}{(H+H_{res})^2 + \Delta H^2}\right\} \label{dyson}
\end{equation}
This is an asymmetric Lorentzian line, which includes both
absorption and dispersion, where $\alpha$ denotes the
dispersion-to-absorption ratio. Such asymmetric line shapes are
usually observed in metals, where the skin effect drives electric
and magnetic microwave components out of phase in the sample and
therefore leads to an admixture of dispersion into the absorption
spectra. For samples small compared to the skin depth one expects
a symmetric absorption spectrum ($\alpha$ = 0), whereas for
samples large compared to the skin depth absorption and dispersion
are of equal strength yielding an asymmetric resonance line
($\alpha$ = 1). At high Y concentrations the spectra are nearly
symmetric with respect to the resonance field in accordance with
the pure absorption spectra for $\alpha$=0. With decreasing $x$
they become more and more asymmetric corresponding to an
increasing parameter $\alpha$. To check, whether the skin effect
is the reason for the asymmetric line shape, we have to estimate
the skin depth $\delta = (\rho / \mu_0 \omega)^{0.5}$ from the
electric resistance $\rho$ and the microwave frequency $\omega = 2
\pi \times 9$ GHz ($\mu_0 = 4 \pi \times 10^{-7}$ Vs/Am). Using
the resistivity values determined by Mallik \textit{et
al.~}\cite{Mallik98a} - for example at 60 K for $x = 0$, $\rho$ =
0.25 m$\Omega$cm - we find a skin depth $\delta =$ 5.7 $\mu$m.
Therefore the skin depth is comparable to or even smaller than the
grain size $(\approx 40 \mu m)$ in accordance with the asymmetry
of the ESR spectra. The large increase of the linewidth on
decreasing temperatures (see Fig.~\ref{spectra}, right column)
results in an increasing uncertainty in determining both,
resonance field $H_{\rm res}$ and dispersion-to-absorption ratio
$\alpha$ accurately, because both quantities are correlated, while
the linewidth remains relatively unaffected. As the resistivity of
the present compounds $T \leq 150$ K changes by only 20\%
\cite{Mallik98a}, it is reasonable to fix the parameter $\alpha$
for $T < 50$ K at a value determined for $T>50$ K, where $\alpha$
is found to be nearly temperature independent (e.g.~$\alpha = 0.7$
for $x = 0.2$).
\begin{figure}[ht]
\centering
\includegraphics[width=80mm,clip,angle=-90]{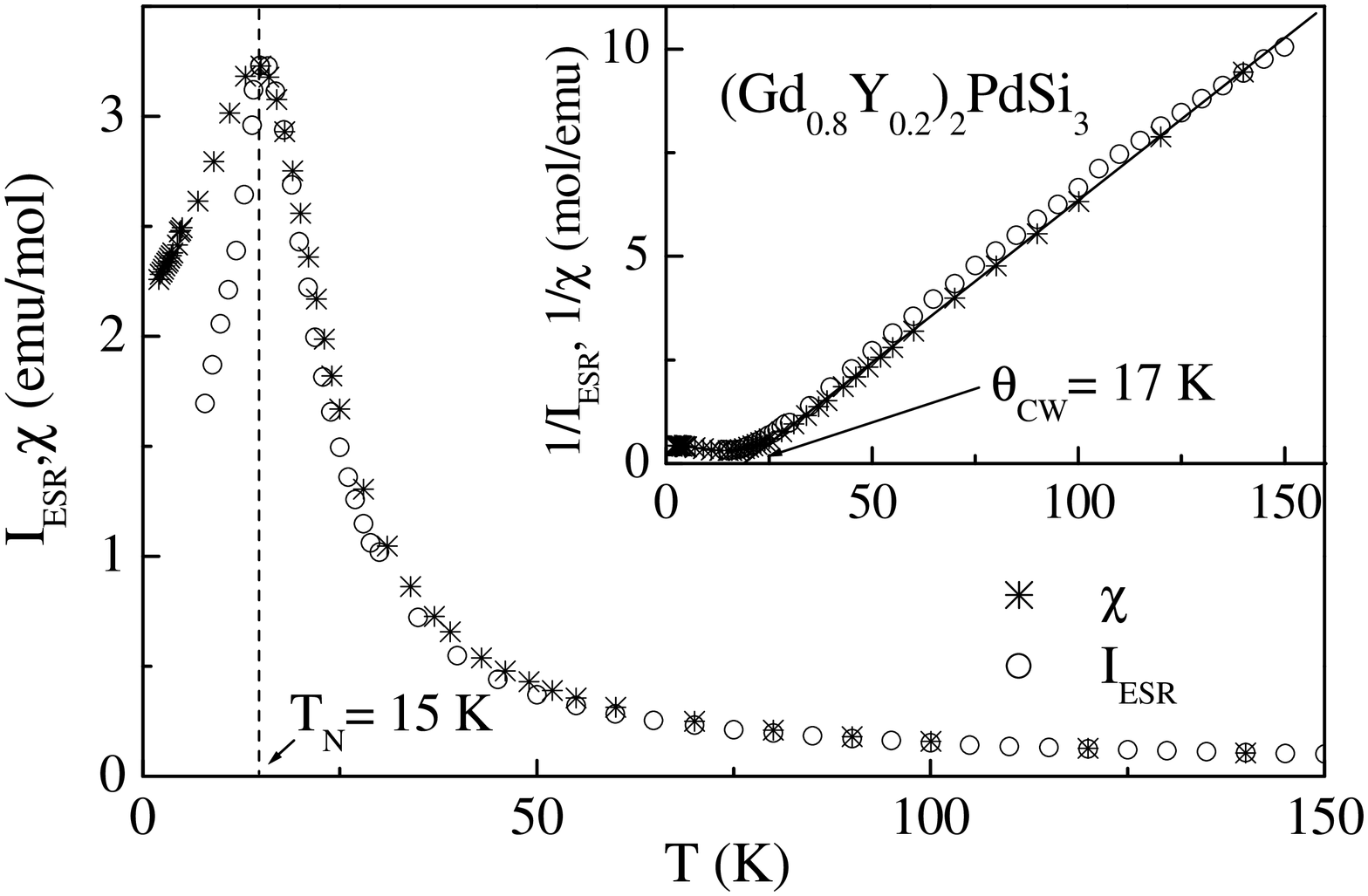}
\vspace{2mm} \caption[]{\label{intensity} Temperature dependence
of the ESR intensity $I_{\mathrm{ESR}}$ and the magnetic
susceptibility $\chi$ (taken from Ref.~5) and the corresponding
reciprocal values (inset) in
(Gd$_{\mathrm{1-x}}$Y$_{\mathrm{x}})_2$PdSi$_3$. The solid lines
in the inset represent fits using a CW behavior.}
\end{figure}
The integrated intensity $I_{\mathrm{ESR}}$ of the resonance line
measures the spin susceptibility $\chi_{S}$ of the ESR probe,
$I_{\mathrm{ESR}}\sim\chi_{S}$. In the case of nonzero dispersion
the intensity is determined by $I_{\mathrm{ESR}} = A \cdot \Delta
H^2 (1+\alpha^2)^{0.5}$, where $A$ denotes the amplitude of
$dP/dH$ and $(1+\alpha^2)^{0.5}$ takes changes of the dispersion
into account. However, one has to be careful with this formula on
approaching $\alpha=1$. If the skin depth is small compared to the
grain size, the microwave probes only a small fraction of the
sample and changes of the resistivity with temperature also change
this fraction. Then the ESR intensity is not a sensible measure of
the spin susceptibility anymore. Fortunately, the relative changes
of the resistivity are small in the temperature regime of
interest, and hence the ESR intensity is even useful, when
$\alpha$ approaches unity. Figure \ref{intensity} shows both
$I_{\mathrm{ESR}}$ and the susceptibility data from Mallik
\textit{et al.}~\cite{Mallik98a} for $x=0.2$. Good agreement
between the two sets of data can be observed throughout the whole
concentration range in the paramagnetic regime. The peaks in the
intensity appear at $T=20,15$ and 8K for $x=0,0.2$ and 0.5
respectively.
\begin{figure}[ht]
\centering
\includegraphics[width=80mm,clip,angle=0]{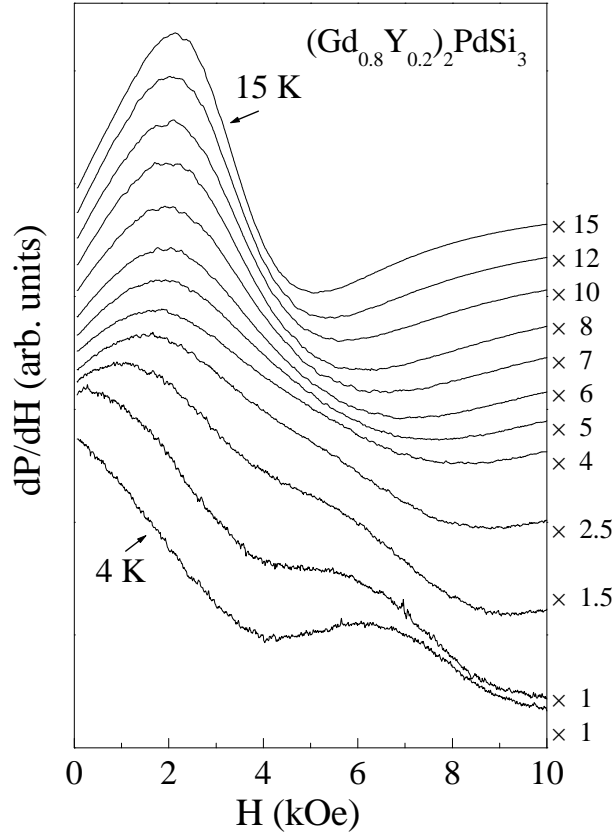}
\vspace{2mm} \caption[]{\label{spectra04} ESR spectra of
(Gd$_{0.8}$Y$_{0.2})_2$PdSi$_3$ for $T\leq T_{\mathrm N}$ (4-15
K), which have to be multiplied by the given amplification factors
in order to get the original amplitude.}
\end{figure}
Below the N\'{e}el-Temperature $T_{\rm N}$ we still observe a
distorted resonance signal for $x\leq 0.5$ as shown in
Fig.~\ref{spectra04} for $x=0.2$, but due to the reasons
mentionend above deviations between ESR intensity and the magnetic
susceptibility appear. The inverse quantities $1/I_{\mathrm{ESR}}$
and $1/\chi$, which are sketched for $x=0.2$ in the inset of
Fig.~\ref{intensity}, exhibit a linear temperature dependence in
the paramagnetic regime ($T>T_{\rm N}$). The samples with
$x=0,0.2$ and 0.5 follow a ($T-\Theta_{\mathrm CW})^{-1}$
Curie-Weiss (CW) law, with positive Curie-Weiss temperatures
$\Theta _{\mathrm CW}$ of 25,17 and 6 K, respectively. For $x>0.5$
the reciprocal ESR intensities $1/I_{\mathrm{ESR}}$ show no
deviations from a Curie law ($\Theta _{\mathrm CW}=0$). The
resonance lines below $T_{\rm N}$ (Fig.~\ref{spectra04}) exhibit a
splitting into a broad absorption band with decreasing
temperature, which can be interpreted in terms of
antiferromagnetic powder spectra with a small antiferromagnetic
gap \cite{Smirnov02}. However, the observation of resonance lines
below $T_{\rm N}$ is a rather unusual feature of intermetallics
\cite{Taylor74}.

\subsection{Temperature dependence of the ESR linewidth and the
resonance field}

\begin{figure}[ht]
\centering
\includegraphics[width=80mm,clip,angle=0]{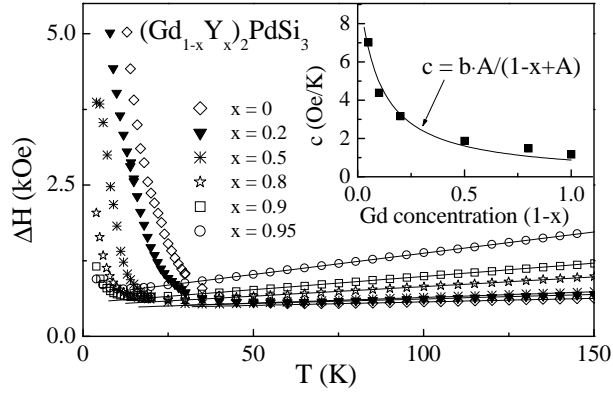}
\vspace{2mm} \caption[]{\label{linewidth} Temperature dependence
of the ESR linewidth $\Delta H$ in
(Gd$_{\mathrm{1-x}}$Y$_{\mathrm{x}})_2$PdSi$_3$. The solid lines
represent a linear fit for $T>T_{min}$, yielding a concentration
dependence (inset) of the slope according to
eq.~(\ref{korringaslope}).}
\end{figure}

Generally, the ESR linewidth $\Delta H$ measures the spin-spin
relaxation rate $1/T_2$ of the Gd-spins, whereas the resonance
field $H_{\mathrm{res}}$ or effective g-value ($\bar
\omega_{\mathrm L}=g\mu_B H_{\mathrm{res}}$) gives information
about the local static magnetic field at the Gd site. The
linewidth is plotted versus temperature for all investigated
samples in Fig.~\ref{linewidth}. As the temperature is lowered,
the linewidth shows a monotonous behaviour that can be fitted very
well with a linear function $\Delta H = c\cdot T$. Then the
linewidth passes a mininum at $T_{min}\approx 2 T_N$ and increases
drastically while approaching the transition temperature to the
magnetically ordered state. This feature is clearly seen for
$x\leq 0.5$, whereas for higher Y concentrations we can only
detect the onset of a line broadening in the temperature range
investigated.

\begin{figure}[ht]
\centering
\includegraphics[width=90mm,clip,angle=-90]{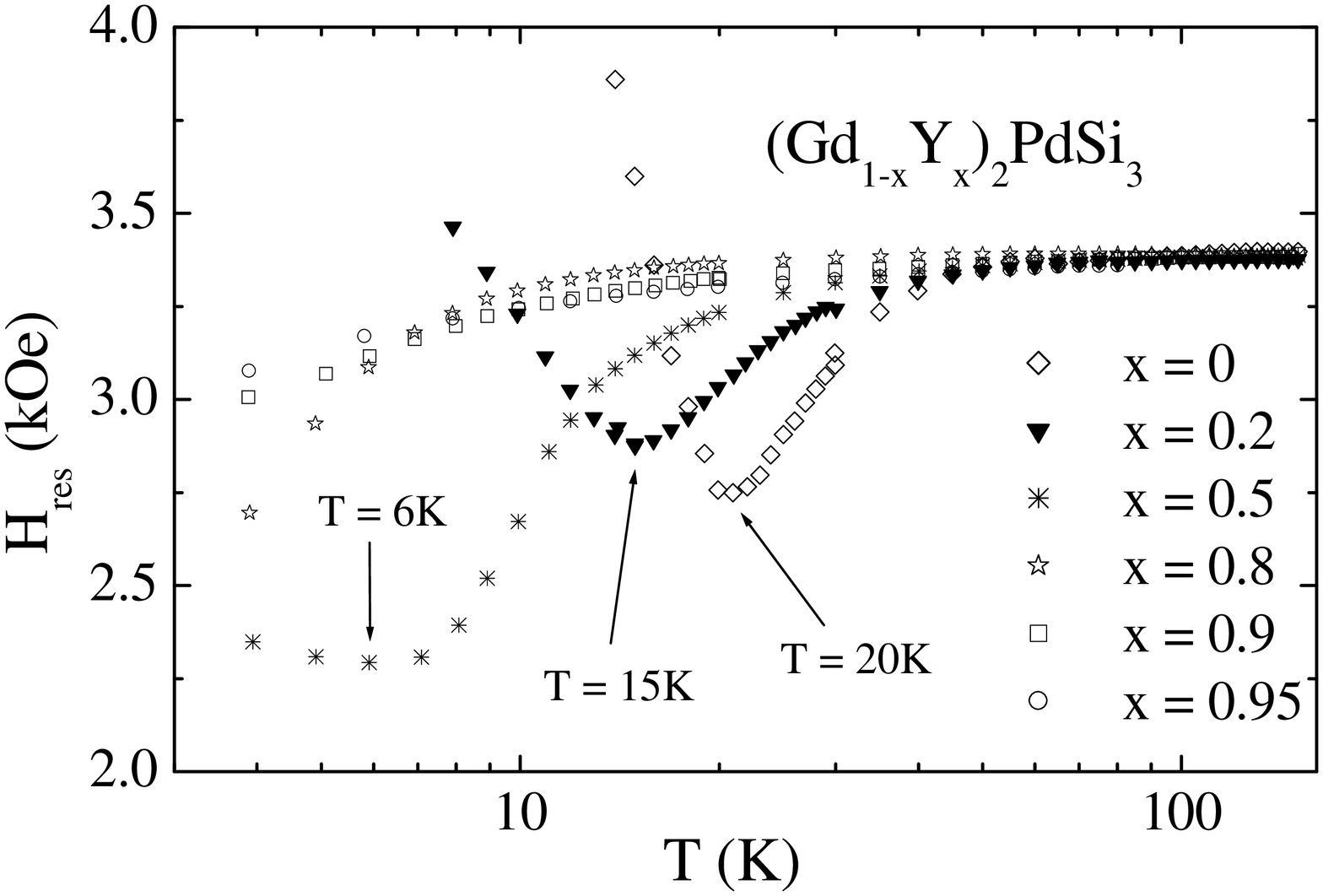}
\vspace{2mm} \caption[]{\label{resonancefield} Temperature
dependence of the ESR resonance field $H_{res}$ in
(Gd$_{\mathrm{1-x}}$Y$_{\mathrm{x}})_2$PdSi$_3$.}
\end{figure}

In Fig.~\ref{resonancefield} we show the temperature dependence of
the resonance field $H_{\mathrm{res}}$ for all samples under
investigation. Again, we have to distinguish between the Gd-rich
concentration range $x\leq 0.5$ and the samples with $x\geq
0.8$:\\ At high temperatures $H_{\mathrm{res}}$ is nearly constant
for all concentrations and yields a g-value $g \approx 1.99$
slightly below the free-electron value. For $T< 100$ K and $x\leq
0.5$ we observe a dramatic shift of the resonance to lower fields,
which coincides with the onset of the line broadening mentioned
above. Finally, a distinct kink and an upturn to higher fields
follows. The peak-like minima for $x=0,0.2,0.5$ are found at 20,
15 and 6 K respectively. These temperatures are consistent with
the ESR intensity and the magnetic-ordering temperatures obtained
from measurements of the heat capacity and the magnetic
susceptibility \cite{Mallik98a}.

\section{Discussion}
The temperature dependencies of the $g$-value and the Gd-ESR
linewidth bear important information on the interaction between
the localized Gd-4f spins and the conduction electrons (for a
review on the theory of ESR in metals, see e.g.~Barnes
\cite{Barnes81}, for experimental reviews see Taylor
\cite{Taylor75} and Elschner and Loidl \cite{Elschner97}). Due to
its half-filled 4f shell ($4f^7$, spin $J=S=7/2$) the orbital
momentum of Gd$^{3+}$ vanishes ($L=0$), and therefore the Gd spin
does not relax directly to the lattice. But the energy is
transferred via the exchange interaction to the conduction
electrons, which themselves couple to the phonons of the lattice.
This process is usually known as Korringa relaxation and yields a
linear increase of the resonance linewidth $\Delta H_{\rm K}
\propto bT$ with increasing temperature $T$, where the slope $b$
is proportional to the squared electronic density of states
$N^2(E_{\rm F})$ at the Fermi energy.

This holds as long as the back-scattering rate $\delta_{ei}$ of
the conduction electrons to the Gd ions is small compared to the
scattering rate $\delta_{el}$ of the conduction electrons to the
lattice. If however the former becomes comparable or even larger
than the latter, the so called bottleneck effect occurs and the
Korringa law is modified by the ratio $B =
\delta_{el}/\delta_{ei}$ of both rates. This results in a reduced
linear increase following \cite{Barnes81}:
\begin{equation}
\label{korringaslope} \Delta H(T)=\Delta H_K(T)\frac{B}{1+B}
\end{equation}
As the back-scattering rate $\delta_{ei}$ is proportional to the
concentration of the localized spins (here $\delta_{ei}\sim$ Gd
concentration $(1-x)$), it is possible to prove the influence of
the bottleneck effect in a given compound by the dependence of the
linewidth on the concentration of magnetic ions.

The inset of Fig.~\ref{linewidth} shows the slope $c$ of the
linear high-temperature increase of the linewidth $\Delta
H(T)=c\cdot T$ in (Gd$_{1-x}$Y$_x$)$_2$PdSi$_3$ as a function of
the Gd-concentration (1-x). It is well fitted by eq.~(2), assuming
a constant electron-lattice relaxation rate $\delta_{el}$, where
$B$ can be expressed as $B=A/(1-x)$ with a constant parameter $A$.
The resulting Korringa rate $b=10$~Oe/K is typical for usual
metals and related intermetallic compounds (e.g. LaCu$_2$Si$_2$
\cite{Schlott88}). Hence, at high temperatures the system behaves
like a regular metal under the strong influence of the bottleneck
effect for all Gd concentrations. Usually in unbottlenecked
metallic hosts the $g$-value of Gd differs from the value in
non-metallic hosts by $\Delta g =N(E_F)J_{ie}(0)$ with the
coupling constant $J_{ie}$ between the Gd-spins and the band
states \cite{Barnes81}. The fact that at high temperatures the $g$
value is not strongly shifted from $g=1.997$, which is usually
observed for Gd$^{3+}$ in insulators \cite{Elschner97}, is in
accordance with the strong bottleneck. The 122-type compounds have
been intensively investigated by ESR by Kaczmarska and coworkers
during the last 15 years \cite{Kaczmarska96}. They found that the
ESR behaviour of dense Gd systems depends on the transition metal
with the thermal broadening parameter $c$ decreasing with
increasing number of $d$ electrons from Co to Cu. For
GdPd$_2$Ge$_2$ a value of 2.0 Oe/K was found,\cite{Kaczmarska89}
which is enhanced in comparison to 1.2 Oe/K for Gd$_2$PdSi$_3$
(see Fig.~3). This result is in agreement with regard to the
relative Gd concentration, which is higher in the latter compound.

\begin{figure}[ht]
\centering
\includegraphics[width=100mm,clip,angle=0]{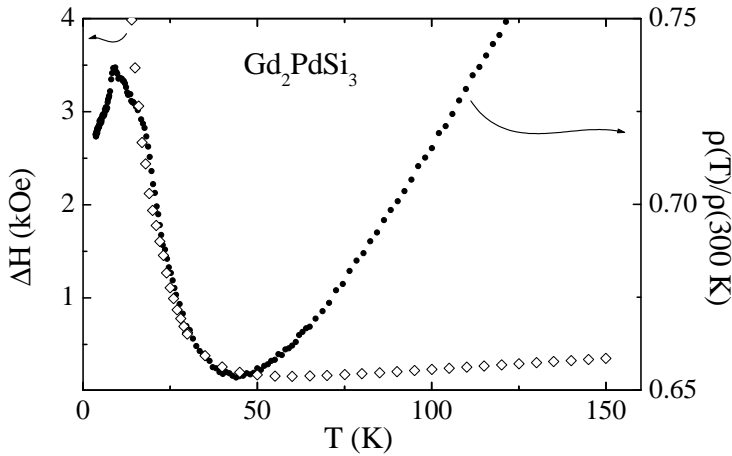}
\vspace{2mm} \caption[]{\label{dHresist} Temperature dependencies
of ESR linewidth ($\diamond$) and resistivity ($\bullet$, taken
from \cite{Mallik98a}) in Gd$_2$PdSi$_{3}$.}
\end{figure}

The increase of the resonance linewidth and the shift of the
resonance field towards lower temperatures on approaching magnetic
order is a common feature of intermetallic compounds containing
rare-earth ions like Gd as reported in the experimental review by
Taylor and Coles \cite{Taylor74}. Generally, for antiferromagnetic
compounds they find no ESR signal below $T_{\mathrm N}$, but a
broadening of the line at about $1.5T_{\mathrm N}<T<10T_{\mathrm
N}$. Some systems like Pd$_3$Gd and EuAl$_4$, however, show an ESR
signal also below $T_{\mathrm N}$, which has been attributed this
as common to materials which undergo metamagnetic transitions in
magnetic fields comparable to the resonance field \cite{Taylor74}.
Exactly such a behaviour has been reported for Gd$_2$PdSi$_3$. At
2 K Saha \textit{et al.}~observed two metamagnetic transitions in
fields of 3 and 9 kOe \cite{Saha99}. On approaching $T_{\mathrm
N}$ from below both transitions coincide at about 3 kOe, the value
of the resonance field. Together with the spectra we observed
below $T_{\mathrm N}$ (see Fig.~\ref{spectra04}) for $x\leq 0.5$,
we conclude that our system belongs to the same peculiar class of
intermetallics as Pd$_3$Gd and EuAl$_4$.\\
Apart from the existence of these signals and the suggested
connection to metamagnetism, we found another very interesting
feature: \\
To analyze the linewidth in (Gd$_{1-x}$Y$_x$)$_2$PdSi$_3$ at low
temperatures in more detail, we remind the magneto-resistance
effects, which indicate the importance of the scattering of the
conduction electrons on the localized Gd spins. Indeed, the
comparison of the resistivity data reported by Mallik et
al.~\cite{Mallik98a} with the EPR linewidth shows that the onset
of the line broadening and the resonance shift
(Fig.~\ref{resonancefield}) coincides with the increase of
resistivity for $x\leq 0.5$. In Fig.~\ref{dHresist} we show the
ESR linewidth and resistivity for the pure compound using a
scaling that reveals the similarity of the increase in resistivity
and linewidth.
Unfortunately, only limited resistivity data has been reported for
EuAl$_4$ \cite{Taylor74} and Pd$_3$Gd making it difficult to judge
unambiguously, whether a correlation between linewidth and
resistivity could be another common property of those materials.
To date, the origin of such a behaviour cannot easily be
explained, but due to the fact that the Gd 4\textit{f} level in
Gd$_2$PdSi$_3$ is about 2 eV below the Fermi level the Kondo
effect cannot be responsible for the observed minima
\cite{Chaika01}, which is in agreement with the fact that the
Korringa rate and therefore the density of states at the Fermi
level is independent on $x$. A comparison with other metallic
systems reveals that also in many ordered and disordered magnetic
compounds a minimum of the ESR linewidth can be found at twice the
critical temperature \cite{Elschner97}. The increase of the
linewidth is attributed to spin fluctuations, which could also
give rise to the resistivity behaviour. Another system where a
correlation between ESR linewidth and resistivity has been
observed is Fe doped La$_{2-x}$Sr$_{x}$CuO$_{4+ \delta}$
\cite{Kruschel93}, where localization effects were considered to
explain that feature. Indeed, a localization scenario has also
been proposed by Mallik \textit{et al.~}in order to explain the
resistivity minimum in (Gd$_{1-x}$Y$_x$)$_2$PdSi$_3$
\cite{Mallik98a}. Very recently, Eremin \textit{et
al.}~\cite{Eremin01} gave an explanation for the
magneto-resistance effects of another Gd alloy, namely GdI$_2$,
which orders ferromagnetically close to room temperature
\cite{Ahn00,Felser99}. The concomitant anomalous peak in the
resistivity has been attributed to scattering of the 5$d$
conduction electrons by localized 4$f^7$ electrons as a result of
the specific topology of the Fermi surface. A similar mechanism
could also account for the correlation in resistivity and
linewidth in (Gd$_{1-x}$Y$_{x}$)$_2$PdSi$_{3}$.

\section{Conclusion}
In conclusion we performed electron-spin resonance measurements in
(Gd$_{1-x}$Y$_x$)$_2$PdSi$_3$. On approaching magnetic order, both
the ESR linewidth and the resonance field are correlated with the
resistivity minimum for $x\leq 0.5$. The high-temperature
behaviour of the linewidth can be well described by a strong
bottleneck effect, yielding a constant density of states at the
Fermi level. The observation of antiferromagnetic-resonance-like
absorption lines below the N\'{e}el temperature can be attributed
to a metamagnetic transition in a magnetic field comparable to the
resonance field. High-Field ESR measurements in single crystals
should be performed in order to investigate the frequency and
field dependence of the resonance lines. Thus, it should be
possible to separate the influences of metamagnetic transitions,
spin fluctuations and localization scenarios. Hopefully, such
investigations will also trigger theoretical studies on
localization effects in compounds with stable localized moments,
which do not show a Kondo effect.

\section{Acknowledgements}
This work was supported in part by the BMBF under contract no.
13N6917 (EKM) and by the Deutsche Forschungsgemeinschaft (DFG) via
the SFB 484.



\end{document}